\documentclass{emulateapj}

\shorttitle{Molecular Gas Destruction by AGN Feedback}
\shortauthors{Schawinski et al.}
\slugcomment{Accepted for publication in the Astrophysical Journal}

\begin{document}

\title{Destruction of Molecular Gas Reservoirs in Early-Type Galaxies by Active Galactic Nucleus Feedback}

\author{
Kevin Schawinski,\altaffilmark{1,2,3}
Chris J. Lintott,\altaffilmark{1}
Daniel Thomas,\altaffilmark{4}
Sugata Kaviraj,\altaffilmark{1}
Serena Viti,\altaffilmark{5}
Joseph Silk,\altaffilmark{1}
Claudia Maraston,\altaffilmark{4}
Marc Sarzi,\altaffilmark{6}
Sukyoung K. Yi,\altaffilmark{7}
Seok-Joo Joo,\altaffilmark{7}
Emanuele Daddi,\altaffilmark{8}
Estelle Bayet,\altaffilmark{3}
Tom Bell\altaffilmark{9}
and Joe Zuntz\altaffilmark{1}
}

\altaffiltext{1}{Department of Physics, University of Oxford, Oxford OX1 3RH, UK.}
\altaffiltext{2}{Department of Physics, Yale University, New Haven, CT 06511, U.S.A.}
\altaffiltext{3}{Yale Center for Astronomy and Astrophysics, Yale University, P.O. Box 208121, New Haven, CT 06520, U.S.A.}
\altaffiltext{4}{Institute of Cosmology \& Gravitation, University of Portsmouth, Portsmouth, PO1 2EG, UK.}
\altaffiltext{5}{Department of Physics and Astronomy, University College London, Gower Street, London WC1E 6BT, UK.}
\altaffiltext{6}{Centre for Astrophysics Research, Science \& Technology Research Institute, University of Hertfordshire, Hatfield AL10 9AB, UK.}
\altaffiltext{7}{Department of Astronomy, Yonsei University, Seoul 120-749, Korea.}
\altaffiltext{8}{Laboratoire AIM, CEA/DSM-CNRS-Universit\'{e} Paris Diderot, DAPNIA/SAp, Orme des Merisiers, 91191 Gif-sur-Yvette, France.}
\altaffiltext{9}{Department of Astronomy, California Institute of Technology, Pasadena, CA 91125, U.S.A.}
\email{kevin.schawinski@yale.edu}

\begin{abstract}
Residual star formation at late times in early-type galaxies and their
progenitors must be suppressed in order to explain the population of
red, passively evolving systems we see today. Likewise, residual or
newly accreted reservoirs of molecular gas that are fuelling star
formation must be destroyed. This suppression of star formation in
early-type galaxies is now commonly attributed to AGN feedback wherein
the reservoir of gas is heated and expelled during a phase of
accretion onto the central supermassive black hole. However, direct
observational evidence for a link between the destruction of this
molecular gas and an AGN phase has been missing so far. We present new
mm-wavelength observations from the IRAM 30m telescope of a sample of
low redshift SDSS early-type galaxies currently undergoing this
process of quenching of late-time star formation. Our observations
show that the disappearance of the molecular gas coincides within less
than $100$ Myr with the onset of accretion onto the black hole and is
too rapid to be due to star formation alone. Since our sample galaxies
are not associated to powerful quasar activity or radio jets, we
conclude that low-luminosity AGN episodes are sufficient to suppress
residual star formation in early-type galaxies. This `suppression
mode' of AGN feedback is very different from the `truncation mode'
linked to powerful quasar activity during early phases of galaxy
formation.
\end{abstract}


\section{Introduction}
The low redshift galaxy population can be divided into blue
starforming and red passively evolving systems
\citep{2004ApJ...600..681B}. The red population is dominated by
elliptical and lenticular (early-type) galaxies, and in order to
explain the properties of these systems as we observe them, almost all
residual star formation in them must be suppressed. This suppression
must take place via the heating or expulsion of the molecular gas
which is the fuel for star formation \citep{1986ApJ...303...39D,
2003ApJ...599...38B} and is thought to be driven by the energy input
from accreting supermassive black holes at the galaxies' centers in a
process called \textit{AGN feedback} \citep{1998A&A...331L...1S,
2003ApJ...599...38B, 2005Natur.433..604D, 2005MNRAS.364..407C,
2005MNRAS.358L..16K, 2006Natur.442..888S, 2007MNRAS.382..960K,
2007arXiv0712.3289K}.

The stellar populations of early-type galaxies indicate that the most
massive galaxies are the first to undergo this process of suppression,
forming their stars earliest and on the shortest time scales
\citep{2005ApJ...621..673T, 2005ApJ...632..137N, 2006AJ....131.1288B,
2007ApJ...669..947J, 2008MNRAS.388...67K, 2008MPLA...23..153K}. These
intense episodes of star formation are followed by passive evolution
with almost no subsequent star formation, although evolution via
processes such as dry mergers is still possible
(e.g. \citealt{2003ApJ...597L.117K, 2005AJ....130.2647V}). It has been
suggested that the mass at which galaxies undergo the transition from
star-forming to quiescent decreases over cosmic time
\citep{2005ApJ...625...23B, 2006ApJ...651..120B}, and so less massive
galaxies have more extended histories of star formation
\citep{2005ApJ...621..673T}.

As new molecular gas is continuously supplied by accretion, mergers
and stellar mass loss, most current models of galaxy formation
suppress residual star formation by invoking a phase of accretion onto
the supermassive black hole at the center of the galaxy
\citep{1998A&A...331L...1S, 2005MNRAS.364..407C, 2006Natur.442..888S,
2006MNRAS.365...11C} as the energy liberated by core-collapse
supernovae is insufficient \citep{1986ApJ...303...39D,
2003ApJ...599...38B}. The resulting active galactic nucleus (AGN)
phase includes jets, radiation and outflows liberating sufficient
energy to destroy the molecular gas reservoir by heating and expelling
it, thus suppressing star formation by depriving the host galaxy of
its fuel. Observations indicate that the mass of galaxies and their
supermassive black holes are correlated \citep{2000ApJ...539L...9F,
2000ApJ...539L..13G}, suggesting that the two co-evolve. The
destruction of the gas reservoir may then in turn terminate the growth
of the black hole as it deprives itself of material for further
accretion as these systems transition to a maintenance mode
suppressing cooling flows due to stellar mass loss and cold accretion
\citep{1997ApJ...487L.105C, 2007ApJ...665.1038C,
2006MNRAS.370..645B}. Alternative -- or perhaps complementary -- gas
heating mechanisms to AGN feedback have been proposed by
\cite{2007MNRAS.380..339B} and
\cite{2008ApJ...680...54K}. Investigating the role of AGN feedback
directly is thus an important observational goal.

 Early-type galaxies are known to pass through episodes of accretion
onto their central black hole, but evidence which links this phase of
their evolution to the removal of the molecular gas reservoir has
until now been lacking. In this Paper, we present observational
evidence for this process occurring in low- to intermediate-mass
early-type galaxies at low redshift and identify low-luminosity AGN as
the culprit.

Throughout this work, we assume cosmological parameters $(\Omega_{m} =
0.3, \Omega_{\Lambda} = 0.7, H_{0} = 70)$, consistent with the
\textit{Wilkinson Microwave Anisotropy Probe} (WMAP) Third Year
results and their combination of results with other data
\citep{2007ApJS..170..377S}.

\section{Observing AGN feedback in action}

\begin{figure*}[!ht]
\begin{center}

\includegraphics[angle=90, width=\textwidth]{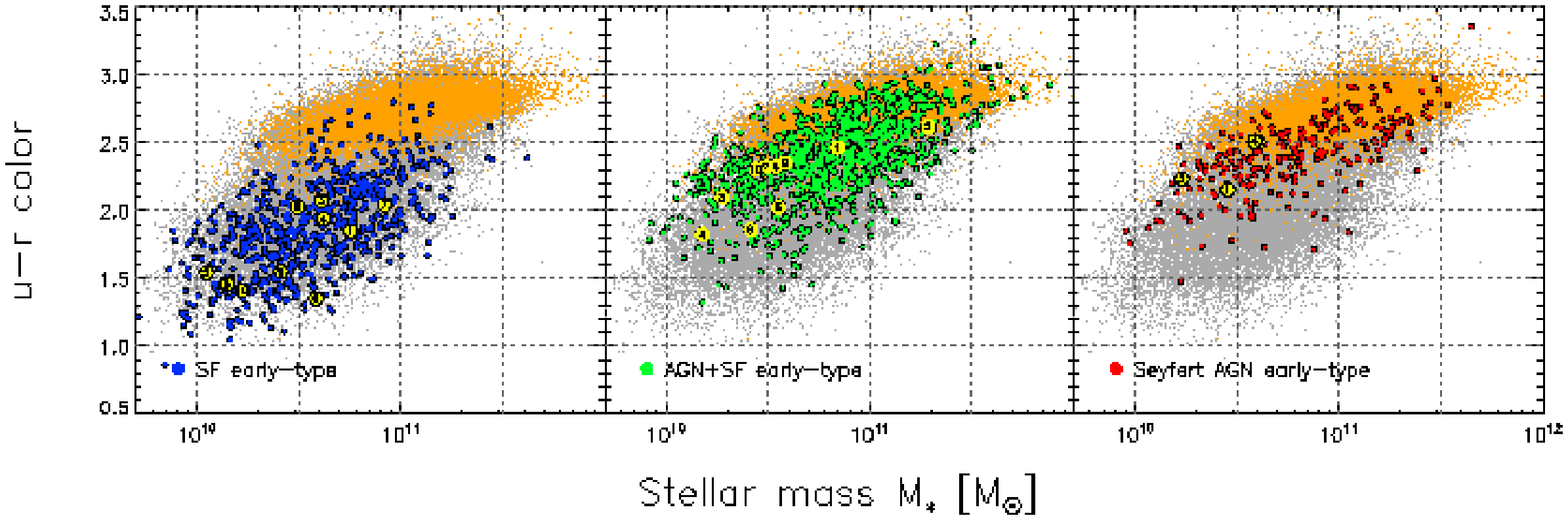}
\vspace{0.05 in}\\
\begin{center}
\includegraphics[angle=90, width=0.28\textwidth]{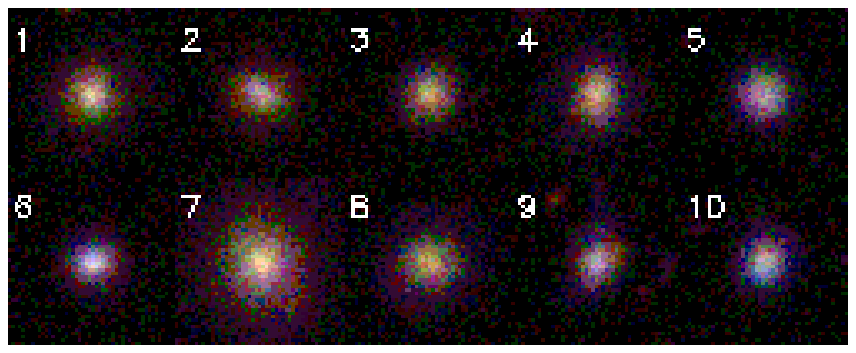}
\includegraphics[angle=90, width=0.28\textwidth]{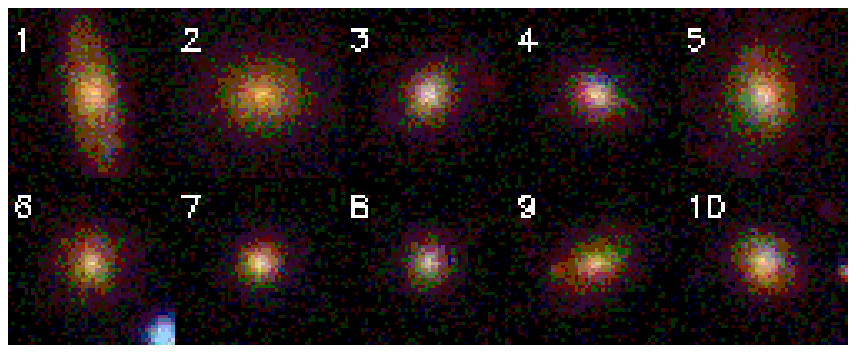}
\includegraphics[angle=90, width=0.28\textwidth]{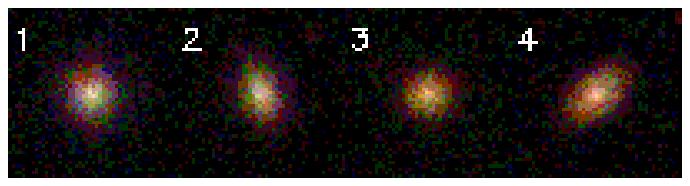}
\end{center}
\caption{The color-mass relationship for the galaxies in our
sample. We plot the galaxy stellar mass derived by spectral energy
distribution fitting using the models of \cite{2005MNRAS.362..799M},
versus the optical $u-r$ color. In each panel, morphological
late-types are gray, quiescent early-types are orange and the various
active early-types (classified by optical emission line ratios) are
colored in in each panel. From left to right: blue (SF), green
(AGN+SF), red (Seyfert AGN). We mark the galaxies observed with the
IRAM 30m telescope with larger points, number them and show their SDSS
$gri$ composite color images \citep{2004PASP..116..133L} below in
three blocks corresponding to the three classifications. This Figure
illustrates the progression of early-type galaxies from the blue cloud
of starforming galaxies via an AGN phase to the red sequence of
passive galaxies. }

\label{fig:sample}

\end{center}
\end{figure*}

\begin{figure*}[!ht]
\begin{center}

\includegraphics[angle=90, width=\textwidth]{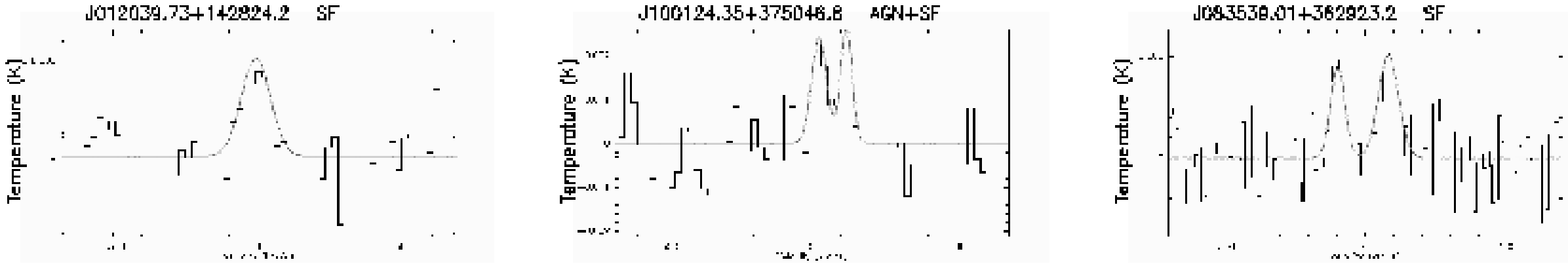}
\includegraphics[angle=90, width=\textwidth]{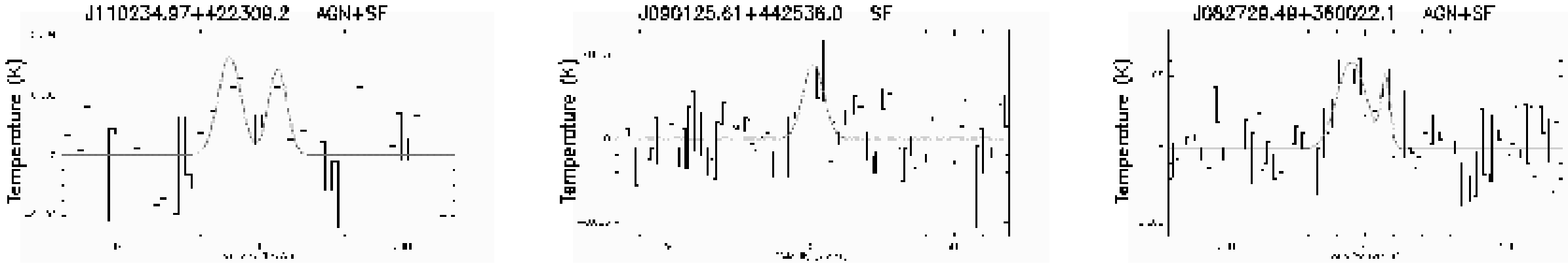}
\includegraphics[angle=90, width=\textwidth]{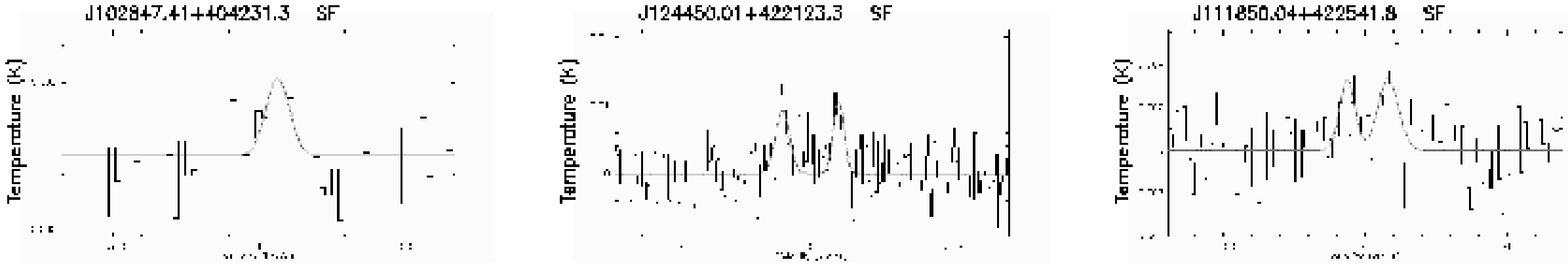}
\includegraphics[angle=90, width=\textwidth]{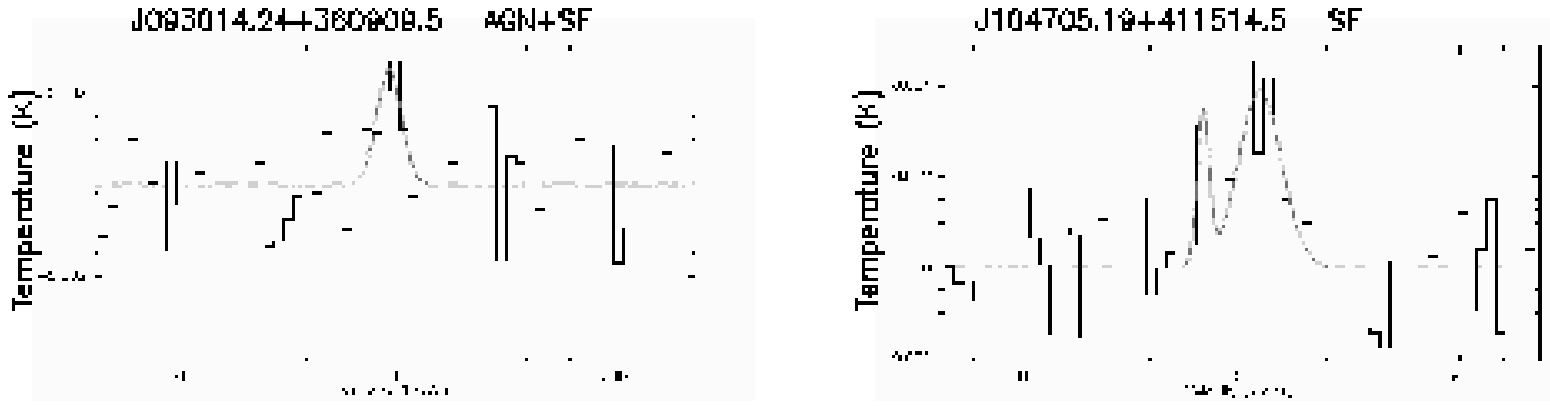}
\vspace{0.1 in}\\
\includegraphics[angle=90, width=\textwidth]{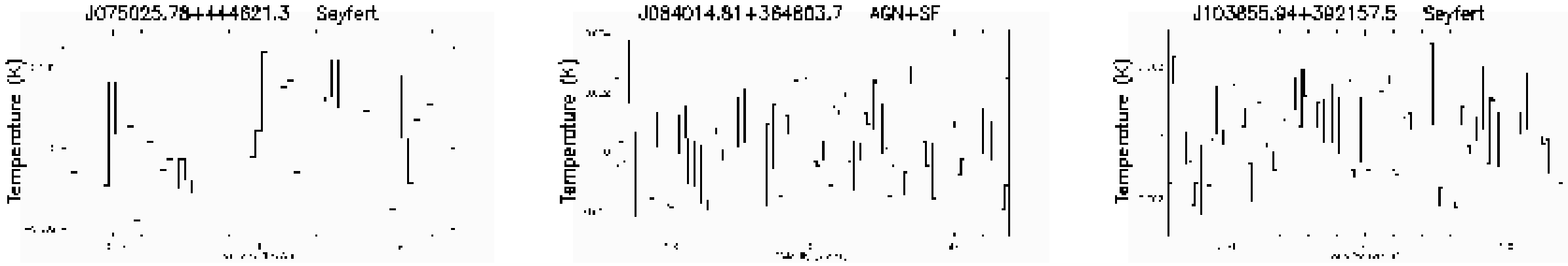}
\caption{CO$(1\rightarrow 0)$ line profiles for our targets. We plot
the temperature as a function of velocity measured from the redshift
of the host galaxy. For each object, we give the SDSS name and
emission line classification (SF, AGN+SF, Seyfert). In the top four
rows, we show the detections, while in the bottom row, we show a
sample of three non-detections. For the detections, we indicate the
best-fit line profile from our \texttt{CLASS} data reduction. Note
that all detected lines are relatively narrow ($\sim70~ \rm kms^{-1}$)
and many objects show double-peaked profiles.}
\label{fig:spectra}

\end{center}
\end{figure*}

\begin{table*}
\caption{SDSS \& IRAM Data}
\tiny
\label{tab:cat}
\begin{tabular}{@{}lrrrrrrrrrl}
\hline
\hline
SDSS Object Name & R.A.    & Dec.    &  Emission      & Stellar Mass                   & Total CO Flux$^2$            &  Molecular Gas            & Age of            &\\
                 &         &         &  Line Class$^1$&  $\rm M_{\rm stellar}$             & $\rm S_{\rm CO(1\rightarrow 0)}$ &  Mass$^3$ $M_{\rm gas}$   & Starburst$^4$ $t_{y}$&\\
                 & (J2000) & (J2000) &                & $(\times 10^{10}~M_{\odot})$     & $\rm(K~kms^{-1})$                 & $(\times 10^{8}~M_{\odot})$ & $(Myr)$             &\\
\hline
SDSS J075109.77+342636.5 &     07 51 09.8 &+34 26 36.6 &      SF &    $5.7 \pm 1.03$ & $<0.07$          &        $ <1.88$ &      $180^{70}_{-70} $ \\
SDSS J012039.73+142824.2 &     01 20 39.7 &+14 28 24.2 &      SF &    $1.1 \pm 0.07$ & $0.63\pm 0.19$   &  $6.0 \pm 1.89$ &        $70^{40}_{-40}$ \\
SDSS J002227.39+135430.6 &     00 22 27.4 &+13 54 30.6 &      SF &    $4.1 \pm 0.31$ & $<0.07$          &        $ <2.03$ &      $340^{70}_{-60} $ \\
SDSS J011458.35+142726.3 &     01 14 58.4 &+14 27 26.4 &  AGN+SF &    $7.0 \pm 1.12$ & $<0.14$          &        $ <2.90$ &      $540^{80}_{-80} $ \\
SDSS J000336.67+141433.9 &     00 03 36.7 &+14 14 34.0 &  AGN+SF &   $19.3 \pm 1.67$ & $<0.14$          &        $ <5.26$ & $1990^{1480}_{-1720} $ \\
SDSS J100124.35+375046.8 &     10 01 24.4 &+37 50 46.9 &  AGN+SF &    $1.9 \pm 0.28$ & $0.28\pm 0.055$  &  $2.5 \pm 0.49$ &     $460^{200}_{-200}$ \\
SDSS J083539.01+362923.2 &     08 35 39.0 &+36 29 23.2 &      SF &    $4.2 \pm 0.43$ & $0.73\pm 0.10$   &  $7.5 \pm 1.04$ &       $200^{70}_{-70}$ \\
SDSS J110234.97+422309.2 &     11 02 35.0 &+42 23 09.2 &  AGN+SF &    $1.5 \pm 0.27$ & $0.56\pm 0.11$   &  $6.5 \pm 1.35$ &       $150^{80}_{-60}$ \\
SDSS J090125.61+442536.0 &     09 01 25.6 &+44 25 36.0 &      SF &    $1.4 \pm 0.25$ & $0.39\pm 0.070$  &  $3.8 \pm 0.68$ &       $190^{40}_{-40}$ \\
SDSS J082729.49+360022.1 &     08 27 29.5 &+36 00 22.1 &  AGN+SF &    $3.5 \pm 0.59$ & $0.97\pm 0.16$   &  $9.6 \pm 1.66$ &      $150^{100}_{-90}$ \\
SDSS J082104.65+412108.5 &     08 21 04.7 &+41 21 08.6 &  AGN+SF &    $3.5 \pm 0.63$ & $<0.07$          &        $ <1.65$ &    $310^{190}_{-130} $ \\
SDSS J102847.41+404231.3 &     10 28 47.4 &+40 42 31.4 &      SF &    $3.8 \pm 1.11$ & $0.53\pm 0.1$    &  $5.3 \pm 1.00$ &       $190^{40}_{-40}$ \\
SDSS J124450.01+422123.3 &     12 44 50.0 &+42 21 23.4 &      SF &    $8.3 \pm 1.69$ & $1.38\pm 0.190$  & $13.4 \pm 1.84$ &      $140^{100}_{-90}$ \\
SDSS J125716.71+424625.9 &     12 57 16.7 &+42 46 26.0 &      SF &    $3.1 \pm 0.19$ & $<0.21$          &        $ <4.09$ &    $310^{120}_{-110} $ \\
SDSS J075025.78+444621.3 &     07 50 25.8 &+44 46 21.3 & Seyfert &    $2.8 \pm 0.31$ & $<0.07$          &        $ <1.18$ &    $470^{150}_{-140} $ \\
SDSS J111850.04+422541.8 &     11 18 50.0 &+42 25 41.8 &      SF &    $2.6 \pm 0.47$ & $0.55\pm 0.10$   &  $6.2 \pm 1.13$ &       $170^{50}_{-50}$ \\
SDSS J084014.81+364803.7 &     08 40 14.8 &+36 48 03.7 &  AGN+SF &    $3.1 \pm 0.58$ & $<0.07$          &        $ <1.45$ &    $900^{460}_{-500} $ \\
SDSS J103855.94+392157.5 &     10 38 55.9 &+39 21 57.6 & Seyfert &    $1.7 \pm 0.24$ & $<0.07$          &        $ <1.40$ &    $960^{330}_{-380} $ \\
SDSS J093014.24+360909.5 &     09 30 14.2 &+36 09 09.6 &  AGN+SF &    $2.6 \pm 0.53$ & $0.17\pm 0.065$  &  $2.1 \pm 0.79$ &       $100^{70}_{-80}$ \\
SDSS J082115.75+355924.6 &     08 21 15.8 &+35 59 24.7 & Seyfert &    $4.0 \pm 0.41$ & $<0.07$          &        $ <1.68$ &   $1080^{660}_{-430} $ \\
SDSS J082507.10+410319.4 &     08 25 07.1 &+41 03 19.5 &  AGN+SF &    $3.8 \pm 0.60$ & $<0.07$          &        $ <1.51$ &    $220^{170}_{-120} $ \\
SDSS J084216.23+360141.4 &     08 42 16.2 &+36 01 41.5 & Seyfert &    $3.8 \pm 0.35$ & $<0.07$          &        $ <1.37$ & $1740^{1020}_{-1150} $ \\
SDSS J104705.19+411514.5 &     10 47 05.2 &+41 15 14.5 &      SF &    $1.7 \pm 0.31$ & $0.34\pm 0.090$  &  $3.4 \pm 0.89$ &       $190^{40}_{-40}$ \\
SDSS J103843.56+404221.0 &     10 38 43.6 &+40 42 21.0 &  AGN+SF &    $2.8 \pm 0.49$ & $<0.07$          &        $ <1.39$ &    $890^{380}_{-440} $ \\
\hline
\end{tabular}
\begin{flushleft}
\tiny
$^1$ BPT emission line classification, described in detail in Schawinski et al. (2007).\\
$^2$ Upper limits for the total CO flux are based on the rms noise of the smoothed spectrum and assuming a line of width $70~kms^{-1}$, they are $1\sigma$ limits.\\
$^3$ Molecular gas masses are derived from the total CO luminosity and assuming a conversion from CO to molecular gas mass of $\alpha = 1.5~ \rm K~km~s^{-1}~pc^2$ (Evans et al. 2005). The gas mass upper limits are $2\sigma$ as in the Figures.\\
$^4$ The age of the last episode of star formation $t_{y}$ is taken from Schawinski et al. (2007) and is based on fitting the UV/optical/NIR spectral energy distribution and the stellar (Lick) absorption indices.
\end{flushleft}
\end{table*}

\begin{table}
\caption{IRAM Data Line Properties$^1$}
\tiny
\label{tab:cat2}
\begin{tabular}{@{}lrrrrrrrrrl}
\hline
\hline
SDSS Object Name$^2$ & Line Width & Line Peak& \\
                 & $\rm kms^{-1}$ & $\rm K$     & \\     
\hline
SDSS J012039.73+142824.2 & $120\pm 20$ & $5.1 \times 10^{-3}$ & \\
SDSS J100124.35+375046.8 & $60\pm 25$  & $2.4 \times 10^{-3}$ & \\
"                        & $35\pm 12$  & $3.3 \times 10^{-3}$ & \\
SDSS J083539.01+362923.2 & $55\pm 18$  & $4.5 \times 10^{-3}$ & \\
"                        & $80\pm 17$  & $5.1 \times 10^{-3}$ & \\
SDSS J110234.97+422309.2 & $87\pm 25$  & $3.2 \times 10^{-3}$ & \\
"                        & $71\pm 28$  & $2.8 \times 10^{-3}$ & \\
SDSS J090125.61+442536.0 & $85\pm 15$  & $4.5 \times 10^{-3}$ & \\
SDSS J082729.49+360022.1 & $120\pm 22$ & $5.9 \times 10^{-3}$ & \\
"                        & $37\pm 25$  & $4.7 \times 10^{-3}$ & \\
SDSS J102847.41+404231.3 & $60\pm 28$  & $1.7 \times 10^{-3}$ & \\
"                        & $96\pm 43$  & $2.2 \times 10^{-3}$ & \\
SDSS J124450.01+422123.3 & $120\pm 22$ & $6.4 \times 10^{-3}$ & \\ 
"                        & $43\pm 7 $  & $1.0 \times 10^{-3}$ & \\
SDSS J111850.04+422541.8 & $20\pm 8 $  & $4.4 \times 10^{-3}$ & \\
"                        & $260\pm 60 $& $2.1 \times 10^{-3}$ & \\
SDSS J093014.24+360909.5 & $65\pm 30 $ & $2.5 \times 10^{-3}$ & \\
SDSS J104705.19+411514.5 & $18\pm 10 $ & $5.4 \times 10^{-3}$ & \\
"                        & $116\pm 26 $& $3.9 \times 10^{-3}$ & \\
\hline
\end{tabular}
\begin{flushleft}
\tiny $^1$ Only for those with CO$(1\rightarrow 0)$ detections.\\  
$^2$
Objects with two components are listed twice with each component line
profile parameters separately.
\end{flushleft}
\end{table}

\begin{figure}[!h]
\begin{center}

\includegraphics[angle=90, width=0.49\textwidth]{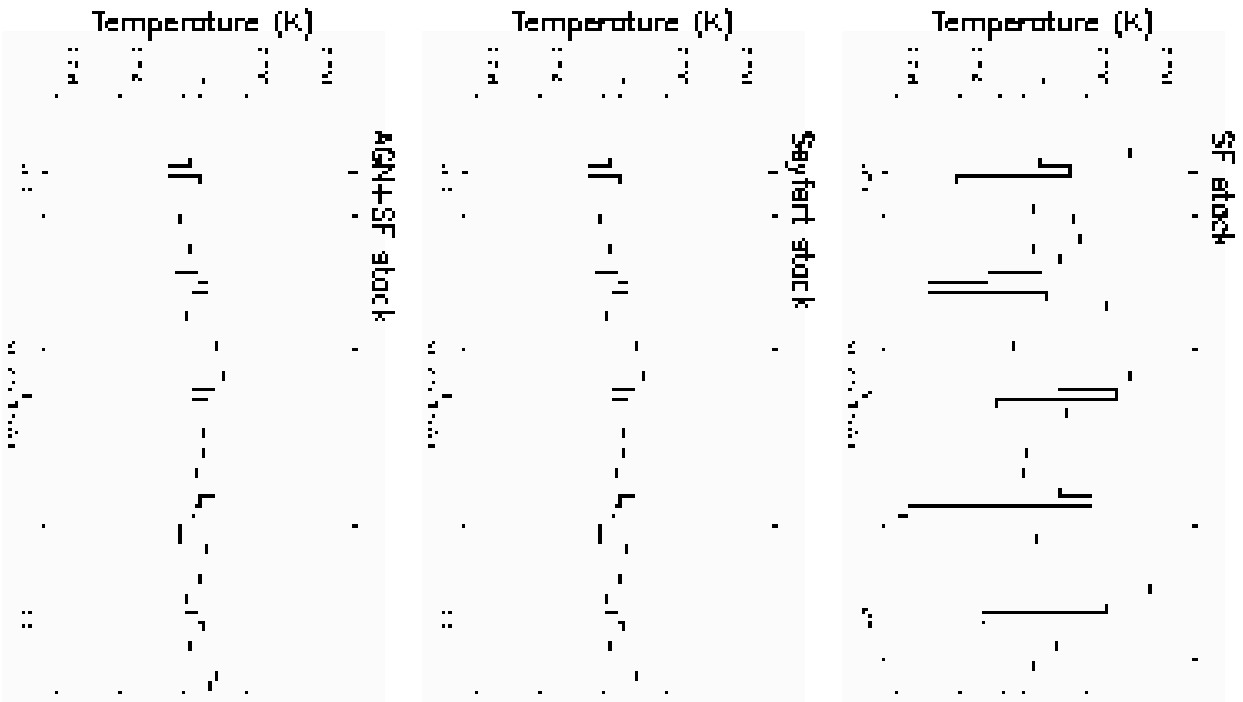}
\caption{CO$(1\rightarrow 0)$ stacked spectra for the Seyferts and the
AGN+SF non-detections. Assuming a very broad line as a worst-case
scenario with a FWHM of $300~ \rm kms^{-1}$, the 2$\sigma$ gas mass
upper limit of the SF, SF+AGN and Seyfert stacks are $10.0$, $3.4$ and
$1.7 \times 10^{8}~ M_{\odot}$.}
\label{fig:nd_spectra}

\end{center}
\end{figure}

We aim to directly observe the effect of AGN on a galaxy's molecular
gas as it changes from a starforming system to a passive early-type
galaxy. \cite{2007MNRAS.382.1415S}, hereafter S07, have assembled a
sample of approximately 16,000 early-type galaxies from the Sloan
Digital Sky Survey (SDSS; \citealt{2000AJ....120.1579Y}) selected by
visual inspection in the redshift range of $0.05 < z < 0.10$. Visual
inspection of SDSS images is a powerful method
\citep{2004ApJ...601L.127F, 2007AJ....134..579F, 2008arXiv0804.4483L}
and ensures the inclusion of early-type galaxies that are still
experiencing star formation or host AGN \citep{2005ApJ...619L.111Y,
2007ApJS..173..512S, 2007ApJS..173..619K}. We identify the dominant
sources of ionising radiation as star formation (SF), AGN, or both in
each galaxy using ratios of nebular emission lines
\citep{1981PASP...93....5B, 2003ApJ...597..142M,
2006MNRAS.372..961K}. We fit the stellar populations using models
\citep{2003MNRAS.339..897T, 2005MNRAS.362..799M} and quantify the age
and the mass-fraction of the last significant episode of star
formation by combining broadband photometry, from the UV to the
near-infrared, with stellar absorption indices measured from optical
SDSS spectra.

The star formation history parameters used here are taken from S07,
who use a two-burst model for the star formation history. The `old'
burst is modelled as a variable single stellar population (SSP) and
the young burst as an exponentially declining burst with an e-folding
time of 100 Myr. Tests show that even large shifts in the e-folding
time do not generally change the best-fit young age and mass-fraction;
only significantly shorter ($\sim 10$ Myr) or longer ($\sim 1$ Gyr)
e-folding time shift the results and these produce worse fits to the
data. The error bars in the young burst age $t_y$ (e.g. in Figure
\ref{fig:observations}) include both formal errors from the fit to the
photometry and spectral indices and errors introduced by the choice of
parameterization by marginalizing over the other parameters, including
dust and metallicity. The marginalization over the variable `old' age
accounts for any variation of the star formation \textit{prior} to the
most recent event, which is accounted for by the young burst.

S07 find that a large fraction of low- and intermediate-mass
early-type galaxies ($50 < \sigma < 150~ \rm kms^{-1}$; $10^{10} <
M_{\rm stellar} < 10^{11}~ M_{\odot}$) show a clear time sequence
(S07; see also Figure \ref{fig:sample}), which begins with actively
starforming early-type galaxies, which formed typically a few percent
of their stellar mass in the most recent episode of star
formation. This is followed approximately 200 Myr later by a phase in
which both AGN and star formation (AGN+SF) exist together. After the
AGN+SF phase, the luminosity of the AGN increases, while all traces of
star formation disappear. The Seyfert AGN (SY) phase begins
approximately 500 Myr after the beginning of the last episode of star
formation. The AGN then declines in luminosity before the galaxy
eventually settles into passive evolution in terms of its star
formation history, though further evolution via dry mergers may still
occur. This striking coincidence between the aging of stellar
populations and the rapid evolution of the nebular emission from being
excited by star formation to being powered by black hole accretion
suggests a role for the AGN in suppressing star formation. A possible
caveat on the nature of the SF phase is the possibility that an AGN is
present even in the first starforming phase, but is sufficiently
obscured to leave the global ISM unaffected and thus does not
influence the optical emission lines. Whether such an obscured AGN
already suppresses (or perhaps even drives) star formation is unclear
and further observations are necessary to investigate this question.

Similar observational results have also been presented by
\cite{2007ApJS..173..267S}, who discuss a possible evolutionary
sequence from star formation to quiescence via an AGN phase. The key
difference between \cite{2007ApJS..173..267S} and S07 is that S07
consider specifically the early-type galaxy population and that S07
demonstrate that the galaxies along the sequence from star formation
to quiescence via the AGN phase form a genuine evolutionary
sequence. Furthermore, \cite{2007ApJ...671..243G} find that at a fixed
velocity dispersion, early-type galaxies hosting LINERs are
systematically younger than their quiescent counterparts, sampling the
penultimate evolutionary phase of S07. The results of S07 favor an AGN
feedback interpretation, but cannot rule out alternative explanations.

If the AGN are responsible for the destruction of the molecular gas,
this material should disappear abruptly with the appearance of the AGN
in the AGN+SF and Seyfert phase. If instead the star formation in
these blue early-type galaxies suffices to exhaust or heat the gas
reservoir, we should see a steady decline of the amount of molecular
gas with time.

In order to test these competing hypotheses, we measure the molecular
gas content of a sample of galaxies along this time
sequence. Molecular hydrogen is exceedingly difficult to observe
directly, so CO -- which is an excellent tracer of molecular gas mass
\citep{1992ApJ...387L..55S, 1998Natur.395..871N} -- is commonly used
instead.

\section{Observations}

From S07, we select early-type galaxies with stellar masses from
$10^{10}-10^{11}~ M_{\odot}$ and randomly choose 10 SF, 10 AGN+SF and
4 Seyfert AGN with velocity dispersion $\sigma < 120~\rm
km~s^{-1}$. The velocity dispersion limit ensures that we select only targets from  S07 that are of 
the type that take part in our AGN feedback time  sequence. Objects with higher velocity 
dispersions typically have  older stellar populations, and must have successfully 
suppressed  cooling over long (Gyr) timescales via mechanisms such as LINERs,  which are 
common in such populations. This sample of early-type galaxies spans the transformation
from star formation via an AGN phase to quiescence and so provides us
with the ideal laboratory to move from inferring to directly probing
the role of AGN. The difference between selecting galaxies from within
categories in this way and randomly selecting galaxies from the entire
sample is at a level which is much less than 2 sigma (assuming a
binomial distribution) and therefore does not produce a significant
change in our results.

We observed the CO$(1\rightarrow 0)$ transition line in this sample of
galaxies along the time sequence with the IRAM (Institutio de
Radioastronom\'{i}a Milim\'{e}trica) 30m telescope at Pico Veleta in
July and December 2007. The IRAM 30M telescope allowed us to
simultaneously also observe the CO$(2\rightarrow 1)$ transition. We
illustrate our sample selection on the color-mass diagram in Figure
\ref{fig:sample} and also provide SDSS images of our targets. The
spectra were reduced with standard \texttt{CLASS} software and first
order polynomial baselines are removed. In Figure \ref{fig:spectra},
we show the reduced spectra, including the fits to line profiles when
the CO$(1\rightarrow 0)$ transition line is detected. We note that the
gas masses derived from our observations are total gas masses. The
size of the IRAM beam at the frequency of CO$(1\rightarrow 0)$ is
$\sim 17.0\arcsec$, while the effective radii of our galaxies is on
the order of a few arcseconds. We compute the CO luminosity of each
target and convert it to a molecular gas mass using a conversion
constant of $\alpha = 1.5~ \rm K~km~s^{-1}~pc^2$
\citep{2005ApJS..159..197E}. The derived gas masses and further galaxy
properties are in Table \ref{tab:cat}. The properties of the detected
lines are listed in Table \ref{tab:cat2}.

\section{Results}
The typical molecular gas masses of those nearby early-type galaxies
with detections have been estimated to range between $10^6 - 10^7 ~
M_{\odot}$ \citep{2007ApJ...657..232S, 2007MNRAS.377.1795C},
corresponding to very low mass fractions. These previous studies
focused on `typical' early-type galaxies drawn from the red sequence
population (see also e.g.
\citealt{1976ApJ...204..365F,1984ApJ...280..107L}). While the gas
content of \textit{spiral} AGN hosts has been studied in detail
\citep[e.g.][]{1986ApJ...311..142B, 2003A&A...407..485G}, no such
studies exist for early-type galaxies. Our work is the first to
systematically observe gas in early-types systems by mapping the
transition from blue to red via an AGN phase and so to investigate the
origin of the low gas masses in red early-type galaxies. In light of
this, we now consider the results of our observations in both the
stacked and individual analysis. We show the results of our
observations in Figure \ref{fig:observations}. In each case, we plot
the mass of molecular gas as a function of the time elapsed since the
start of the last significant episode of star formation $t_y$ from
S07.

In the case of non-detections, we must make an assumption on the
expected line profile to determine the amount of gas that would lead
to a detection. We divide our analysis into two separate avenues here:
first, we perform a very conservative analysis using stacked spectra
to establish that we see a trend between age/emission line class and
gas fraction. After establishing that a trend exists, we then consider
individual galaxies using less conservative upper limits.

\subsection{Analysis of stacked spectra}
We wish to establish the most conservative possible upper limits on
our non-detections. Despite the low rms noise levels of our spectra,
it is conceivable that a very broad line might lurk just below our
sensitivity limits. In order to guard ourselves against this, we stack
the non-detected spectra in each emission line class to increase our
signal-to-noise and assume the broadest line that can reasonably be
expected, a line with FWHM of $300~ \rm kms^{-1}$. From this, we
obtain 2$\sigma$ gas mass upper limits for the non-detected sources in
each class of $10.0$, $3.4$ and $1.7 \times 10^{8}~ M_{\odot}$ for the
SF, AGN+SF and Seyfert classes. The stacked spectra for the SF, AGN+SF
and Seyferts are shown in Figure \ref{fig:nd_spectra}; they show no
evidence for any lines. For the objects with detected lines, we simply
take the mean of the derived gas masses, as stacking might distort the
line profile and thus the derived masses.

\begin{figure*}
\begin{center}

\includegraphics[angle=90, width=0.49\textwidth]{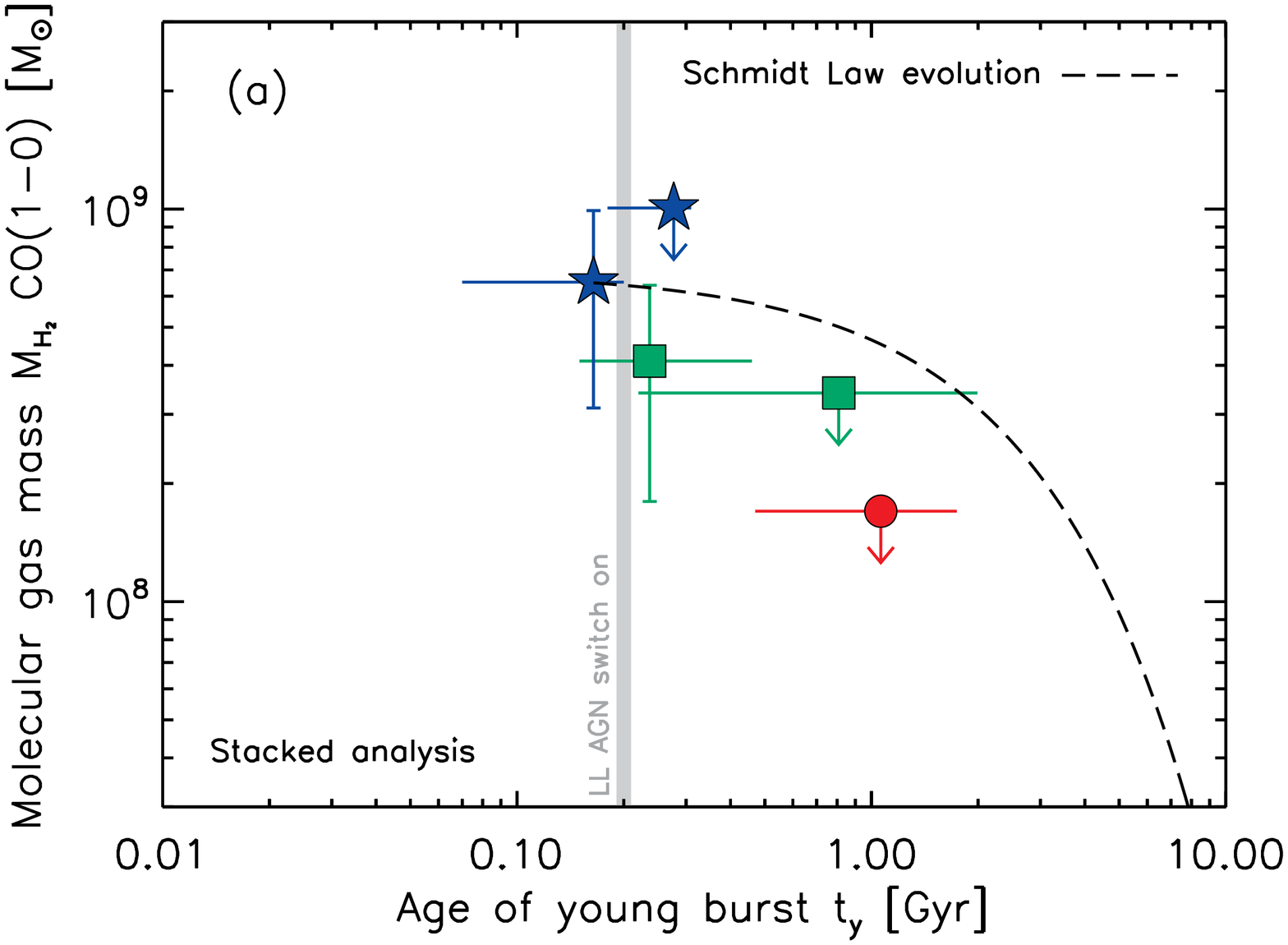}\\
\includegraphics[angle=90, width=0.49\textwidth]{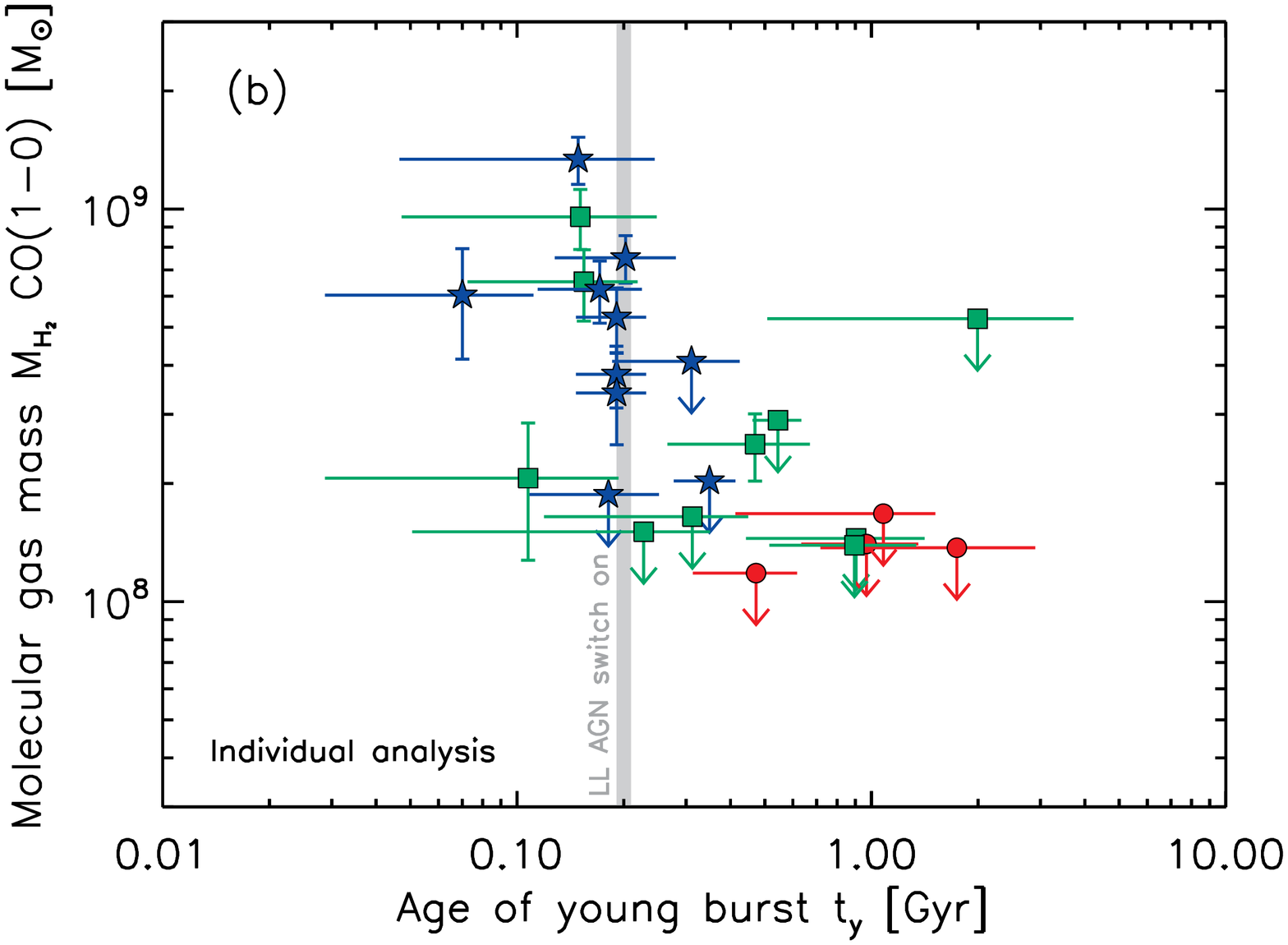}
\includegraphics[angle=90, width=0.49\textwidth]{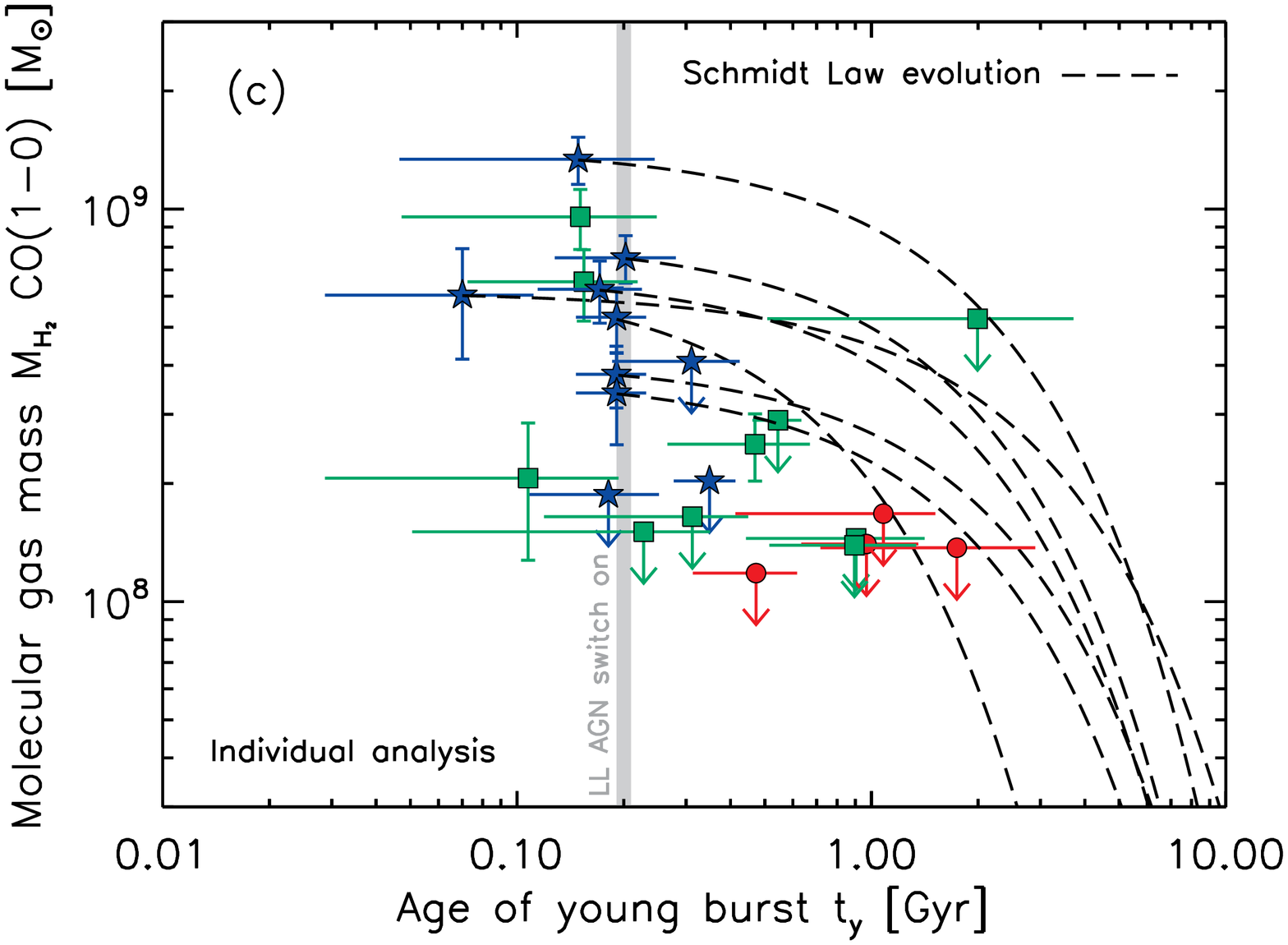}
\caption{The observed molecular gas masses $\rm M_{\rm H_2}$ as a
function of $t_{y}$, the age of the last significant episode of star
formation \citep{2007MNRAS.382.1415S}. The blue stars are early-type
galaxies classified as starforming, the green squares are AGN+SF
composites and the red circles are Seyfert AGN. In the top panel, we
show the results from the analysis of the stacked spectra. The upper
limits are 2$\sigma$ assuming a typical linewidth of 300 $\rm
kms^{-1}$. In the bottom panels, (b) and (c), we show the individual
gas masses for each galaxy. The upper limits are 2$\sigma$ assuming a
typical linewidth of 70 $\rm kms^{-1}$. The values of $t_{y}$ are
based on assuming an e-folding time of 100 Myr for the starburst;
allowing this time to vary simply extends the error bars, but does not
move the best-fit values. Instantaneous bursts or constant star
formation models yield poor fits to the data. In panels (a) and (c),
we show the tracks showing the future evolution of the gas mass
assuming the Schmidt Law for star formation. In (a), we show this
track for the mean SF galaxy, wherease in (c), we show tracks for each
individually detected SF galaxy. These tracks illustrate that given an
evolution driven purely by star formation, the depletion of the gas
reservoirs on the short time scales we observe is not feasible. These
tracks assumes a dynamical time scale of $\tau = 100$ Myr and is
inconsistent with the systematically lower gas masses of the AGN+SF
objects -- whether detected or not. We mark the point in time $t_{y} =
200$ Myr when the low-luminosity AGN in the AGN+SF objects typical
switch on.}
\label{fig:observations}

\end{center}
\end{figure*}

In panel (a) of Figure \ref{fig:observations}, we show the mean gas
masses of those galaxies which are detected in both the SF and the
AGN+SF classes. We furthermore show the upper limits for the stacked
non-detections assuming a very conservative line width of $300~ \rm
kms^{-1}$.  The starforming early-type galaxies at young ages have
high molecular gas masses of the order of up to $\sim 10^{9}~
M_{\odot}$.  On the other hand, the pure (Seyfert) AGN where no traces
of ongoing star formation are left in the optical emission lines are
undetected in CO with upper limits on the gas mass significantly lower
than the younger SF objects. In between lie the AGN+SF objects: the
detections are systematically younger and have mean gas masses
slightly below the SF objects. The non-detected AGN+SF objects are
older than their detected counterparts, but younger than the
Seyferts. During in the AGN+SF phase, at about $200$ Myr after the
onset of star formation, the molecular gas mass drops
precipitously.

It should be noted that the presence of CO correlates with other
measures of star formation and so galaxies with more CO have more
vigorous star formation \citep{1998ApJ...498..541K,
2004ApJ...606..271G, 2005ApJ...630..269N}. However, the star formation
history parameters derived by S07 would account for the implied larger
mass fraction in the young population and still return the correct age
for it.

\subsection{Analysis of individual spectra}
On the basis that we can establish the significance of the observed
trend in panel (a) of Figure \ref{fig:observations} using the stacked
spectra, we now turn to invididual galaxies.  Since S07 resolve the
age of the current episode of star formation and provide emission line
classes for each individual object, we can also plot these to reveal
trends within populations. Half of our detections shows single, narrow
CO$(1\rightarrow 0)$ lines. The other half shows two components. The
median FWHM of the single lines and the individual components is $70~
\rm kms^{-1}$. These relatively moderate widths are consistent with
the fact that the galaxies in our sample are only intermediate-mass
objects and that they are undergoing minor residual star
formation. They are not undergoing major starbursts requiring enormous
gas reservoirs, nor are they spiral galaxies with extended disks with
broad double-horned lines.

In the absence of comparable samples, we use the median FWHM of our
sources with detections of $70~ \rm kms^{-1}$ to estimate the upper
limts on the gas mass in the non-detections. These are not as rigorous
as the limits of the stacked spectra, but as we will show, the results
are consistent. With this analysis, we neglect the possibility of
double-horned profiles in the analysis, which would increase the upper
limits of the non-detections by up to a factor 2. The analysis of the
stacked spectra however does not support this latter scenario.

The individual galaxies show the same trend and allow us some insight
into variations within each population. We present these in panel (b)
of Figure \ref{fig:observations} and the gas masses and ages $t_y$ are
given in Table \ref{tab:cat}. We see here that the SF molecular gas
masses are within a factor of a few of each other and reach up to
$\sim 10^{9}~ M_{\odot}$. All three Seyfert AGN are individually
undetected.  In between are the AGN+SF objects, two of which have high
masses with the remainder either undetected or significantly lower
than the starforming galaxies. Figure \ref{fig:observations}b shows
the same trend as Figure \ref{fig:observations}a. The individual gas
masses and upper limits further support the observation that the gas
mass drops precipitously at $\sim 200$ Myr within the AGN+SF phase.

To assess the significance of the difference between the populations
we divide the sample into two groups, one younger than $200\
\textrm{Myr}$ and one older. We use Bayes' theorem to find the
posterior distribution of the population mean and variance of the two
groups' molecular gas masses by assuming that each group is drawn from
a different normal distribution. Samples on which only upper limits
were found are described as gaussian data points with zero mean and
the appropriate width. We place a flat prior on the population means
and standard deviations. After computing the joint posteriors, we
marginalize over the standard deviation. The result is a probability
distribution for each of the two populations means regardless of their
variances. Together the distributions show that the younger group has a
higher mean with $>95\%$ probability.

\subsection{The case for AGN feedback: the Schmidt Law and gas
  depletion time scales}

\begin{figure*}
\begin{center}

\includegraphics[angle=90, width=\textwidth]{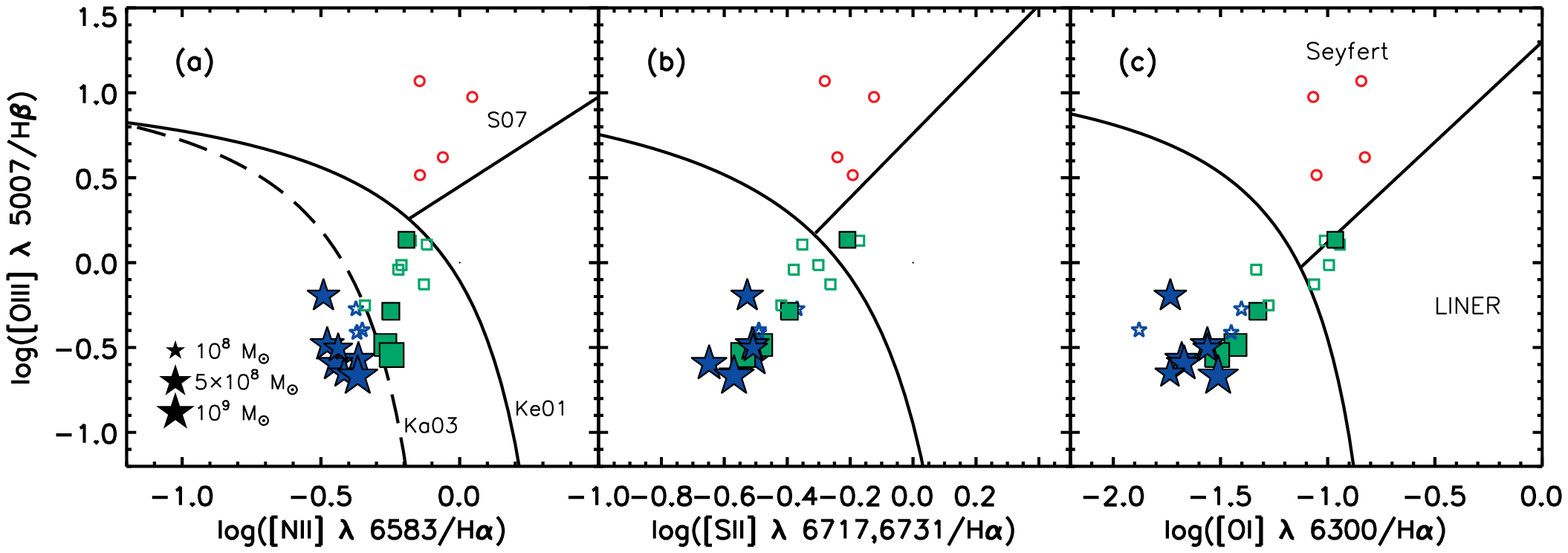}
\caption{The objects in our sample on emission line diagrams
\citep{1981PASP...93....5B, 2001ApJ...556..121K, 2003MNRAS.346.1055K,
2006MNRAS.372..961K}. In panel (a), we delineate starforming galaxies
by the dashed line (Ka03; \citealt{2003MNRAS.346.1055K}). Between the
dashed and solid line (Ke01) reside AGN+SF composites, while beyond
the solid line there are the pure AGN \citep{2001ApJ...556..121K}. The
straight line (S07) divides Seyfert AGN from LINERs. The blue stars
represent starforming early-type galaxies, the green squares AGN+SF
composites and the red circles Seyfert AGN. Filled symbols are
CO$(1\rightarrow 0)$ detections and open symbols are
non-detections. The size of the symbol scales with the logarithm of
the molecular gas mass (see legend in (a)). In panels (b) and (c), we
show a similar line ratio diagram using the classification of
\cite{2006MNRAS.372..961K}. The species [SII] and [OI] used in (b) and
(c) are sensitive to low ionisation states and complement (a) based on
[NII]. Note that the two AGN+SF galaxies with large molecular gas
masses are also modest radio sources.}

\label{fig:bpt}

\end{center}
\end{figure*}

Such a sudden drop in molecular gas mass as we observe cannot be
accounted for by star formation alone. To demonstrate this, we compute
the evolution of the molecular gas mass for both the mean SF galaxy
and each of the SF early-types with CO$(1\rightarrow 0)$ detections,
using the Schmidt Law for star formation \citep{1959ApJ...129..243S,
1998ApJ...498..541K}:

\begin{equation}
SFR = \frac{\epsilon M_{\rm gas}}{t_{\rm dyn}}
  \label{eqno1}
\end{equation}

We assume a fiducial efficiency of 2\% ($\epsilon = 0.02$), which
includes the effect of SN feedback. This law appears to be universally
applicable across galaxy morphologies, including early-type galaxies
\citep{2007MNRAS.377.1795C}. We compute the dynamical times of each
galaxy from its radius and stellar mass ($t_{dyn} =
\sqrt{R^{3}/2GM}$); for the stacked SF galaxies, we use the median
$t_{dyn}$ of 50 Myr of the SF galaxies. We show the resulting
evolutionary tracks in Figure \ref{fig:observations}a and c (dashed
lines).

The depletion time scales derived this way from the Schmidt Law are on
the order of several Gyrs, much longer than the observed timescale for
gas depletion. The tracks shown in Figure \ref{fig:observations}a and
c are inconsistent with a scenario where star formation alone depletes
the gas reservoir.

Would it be possible to adjust the parameters used to result in tracks
for star formation alone that are consistent with our observations?
Adjusting the mass and radius to achieve the rapid gas reservoir
exhaustion we observe results in parameters not characteristic of a
galaxy-wide starburst, but perhaps a circumnuclear
ring. Alternatively, a drastic increase of the star formation
efficiency $\epsilon$ towards unity driven by an AGN
\citep{2005MNRAS.364.1337S} might suffice, though this completes the
circle and brings us back to AGN feedback. Our modeling does
\textit{not} account for the amount of recycled gas from stellar mass
loss or further gas accreted, both of which would further increase the
amount of molecular gas available over time and thus extend the
depletion time scales significantly. This caveat further strengthes
our argument against star formation alone.

This decrease in molecular gas coincides with the time when galaxies
transition from pure star formation to an AGN+SF composite as the AGN
switches on of $\sim 200$ Myr (S07). Figure \ref{fig:bpt} shows this
same trend in the emission line diagnostic diagrams. There the two
AGN+SF composites with the gas masses similar to the starforming
objects reside in close proximity to them; those AGN+SF objects
further away from the starforming locus are either undetected or have
significantly lower gas masses.

\section{Discussion}
Models of hierarchical galaxy formation suggest that AGN feedback
plays an important role in the formation and late-time evolution of
early-type galaxies. Current simulations invoke two modes of black
hole growth and AGN feedback: a `quasar mode'
\citep[e.g.][]{2005MNRAS.361..776S,
2005Natur.433..604D,2008ApJS..175..390H} and a `radio mode'
\citep[e.g.][]{2006MNRAS.365...11C, 2006MNRAS.370..645B,
2007MNRAS.380..877S}.

\subsection{Two modes of AGN feedback}
The quasar mode, implemented in simulations such as that of
\cite{2005MNRAS.361..776S}, assumes that during a major merger a
fraction of the rest mass of the gas accreted onto the central black
hole is liberated as energy injected back isotropically into the gas
reservoir of the host galaxy.  Simulations of gas-rich mergers at very
high redshift \citep{2006ApJ...642L.107N, 2007ApJ...665..187L,
2008ApJS..174...13N} indicate that quasars might be driving massive,
powerful outflows as the energy liberated by black hole accretion is
coupled to the gas in the form of thermal energy. These outflows may
succeed in halting star formation on short time scales
\citep{1992ApJ...391L..53H}, as required for the most massive
early-type galaxies \citep{2005ApJ...621..673T}. On the other hand,
observations of quasars at high redshift yield very high CO
luminosities and hence high inferred gas masses
\citep{2002A&A...387..406C, 2003A&A...409L..47B, 2004ApJ...615L..17W},
which suggests that feedback in these objects does not necessarily
result in the instantaneous removal of molecular gas from the system.

The radio mode (\cite{2006MNRAS.365...11C, 2006MNRAS.370..645B,
2007arXiv0712.3289K}), instead, is important for the evolution of
galaxies and is supposed to occur at later times. A low-level AGN is
thought to be injecting a fraction of the rest mass energy of the hot
gas accreted by the central black hole into the ISM over an extended
period of time to keep the host galaxy quiescent.

\subsection{Molecular gas reservoir destruction by low-luminosity AGN}
Our galaxies help us illuminate the role of AGN in suppressing
late-time star formation. In previous work (S07) we have identified a
sample of low- and intermediate mass early-type galaxies that display
an evolutionary sequence from minor star formation through AGN to the
red sequence, most likely driven by AGN feedback. In the present paper
we investigate through IRAM mm-observations the presence and absence
of cold, molecular gas reservoirs in objects selected along this
sequence.

The results presented here support the presence of late-time AGN
feedback invoked in the simulations. Our observations suggest a
scenario in which molecular gas is removed or heated $\sim 200$ Myr
after the peak of recent star formation in our systems. Most
importantly, this observed drop coincides with the timescale on which
these galaxies transition from being primarily star forming to being
dominated by AGN. We show that significantly longer timescales would
be expected if the molecular gas was to be consumed by star formation
alone.

The AGN detected in our sample are low-luminosity AGN. They are
neither powerful quasars nor radio-loud. Their accretion rates and
efficiencies are low ($10^{6} < L_{\rm [OIII]} < 10^{8} ~L_{\odot}$,
$-3 < \rm log(\it L_{\rm [OIII]}/\sigma^4)\rm < 0$; S07)\footnote{The
quantity $L_{\rm [OIII]}/\sigma^4$, introduced by
\cite{2006MNRAS.372..961K}, is a proxy of the Eddington ratio. $L_{\rm
[OIII]}$ traces the accretion rate, while $\sigma^4$ traces the black
hole mass via the $M_{\bullet}-\sigma$ relation
\citep{2000ApJ...539L..13G, 2000ApJ...539L...9F}.}. Out of our entire
sample only two targets are detected at 1.4 GHz by the Very Large
Array FIRST Survey \citep{1995ApJ...450..559B} at a detection limit of
$3\times 10^{21}~ \rm W~Hz^{-1}$, both of which are AGN+SF objects
with modest $L_{\rm 1.4~GHz}$ of $6.7 \times 10^{21}\rm ~W~Hz^{-1}$
and $2.5 \times 10^{22}\rm ~W~Hz^{-1}$. These two AGN+SF objects are
those with the two highest molecular gas masses.

Some theoretical models have discussed scenarios where low-luminosity,
low accretion efficiency AGN of the kind we see here regulate star
formation \citep{2006ApJS..166....1H,
2007ApJ...665.1038C}. Furthermore, the time delay between peak SF and
AGN Eddington ratio and the moment of `blowout' -- when the AGN drives
out the gas reservoir -- is present in the models of
\cite{2006ApJS..166....1H} and \cite{2006ApJ...639..700H}. The process
we see thus might be physically similar to quasar mode feedback as
envisioned by theorists, but fulfills the task of the radio mode in
moderate-mass early-type galaxies of suppressing further late-time
star formation.

We conclude that low-level AGN appear to be powerful enough to destroy
the molecular gas reservoir and suppress star formation in early-type
galaxies. Furthermore, we note that along the time sequence
established and described by S07, the peak of the AGN accretion rate
and efficiency occur during the Seyfert phase; the AGN phase that
suppresses star formation is less efficient than this peak.

\subsection{Suppression vs Truncation}
At high redshift, early-type galaxies (or their progenitors) experience
vigorous star formation. The large energy output from quasar-like AGN
is required to bring such massive star formation to a halt. We refer
to this type of AGN feedback as `truncation mode'. In order to account
for the observed high star formation efficiency in these galaxies, it
is possible that a `trigger mode' may precede the truncation mode
\citep{2005MNRAS.364.1337S}.

Clearly, the process we are probing with the present work is very
different. We see AGN feedback operating at late times in early-type
galaxies associated with low-luminosity AGN. No quasar-like activity
or powerful radio jets are observed. We witness the suppression of
star formation in early-type galaxies at recent epochs. We therefore
prefer to call this type of AGN feedback `suppression mode' as opposed
to the truncation mode at early times. The low- and intermediate mass
galaxies studied in our undergo very minor episodes of star formation
and thus allow us to study this suppression process. Apparently, the
AGN in this lower mass range is not powerful enough to inhibit
residual star formation completely \citep[see
also][]{2006Natur.442..888S}. As shown in Figure
\ref{fig:observations}, however, without the energy input from AGN,
those episodes of star formation would be more significant and would
drive the objects even further away from the red sequence. The
observed suppression by AGN feedback ensures that these objects will
join or return to passive evolution on the red sequence. Note that the
massive early-type galaxies in our sample, instead, do not show any
signs of residual star formation. As discussed in S07, the LINER
activity detected in some of them might be the `smoking gun' of highly
effective suppression mode AGN feedback.

Finally, it is interesting to note that the suppression mode detected
here acts on timescales around several hundred Myr, which is the
typical timescale of an AGN duty cycle. Hence, suppression of residual
star formation in early-type galaxies at late times may not be a
continuous but a periodic process (see
e.g. \citealt{2007ApJ...665.1038C}).

\section{Conclusions}
The molecular gas reservoirs fuelling late time star formation in
early-type galaxies must be destroyed to account for their passive
evolution on the red sequence. This suppression of star formation in
early-type galaxies is now commonly attributed to AGN feedback wherein
the reservoir of gas is heated and expelled by the energy input from
an accretion phase onto the central supermassive black hole. In this
Paper, we present observational evidence for this process occurring in
low- to intermediate-mass early-type galaxies at low redshift and
identify low-luminosity AGN -- not quasars or radio galaxies -- as the
culprit. The systems studied here are not the massive galaxies
in clusters where the radio mode is generally invoked to suppress
cooling.

In order to empirically determine whether and at which stage AGN are
involved in the destruction of significant molecular gas reservoirs,
we perform an observational investigation. We have observed a sample
of low-redshift intermediate mass early-type galaxies during the
process of moving from the blue cloud to the red sequence via a
low-luminosity AGN phase. We observed galaxies along the time sequence
of S07 with the IRAM 30m telescope near Granada, Spain to determine
the amounts of molecular gas present via the CO$(1\rightarrow 0)$
line.

We perform two separate analyses: First, we take the average gas
masses in each of the SF, SF+AGN and Seyferts galaxies and stack the
spectra of the non-detections to increase signal-to-noise and
therefore the upper limit on the gas mass. For the non-detections, we
assume an extremely conservative line width of $300~ \rm kms^{-1}$ and
find a statistically significant trend. The gas mass in the SF
galaxies drops approximately $200$ Myr after the start of star
formation, during the AGN+SF phase, and is far too rapid to be
accounted for by star formation alone. None of the four Seyfert AGN
were detected, suggesting that the gas reservoirs were already
destroyed in the preceding AGN+SF phase, i.e. prior to the peak AGN
luminosity and accretion efficiency (c.f. S07). The molecular gas
reservoir drops by at least an order of magnitude in mass in a very
short time. The Schmidt law for star formation implies depletion
timescales of several Gyr, even if contributions such as further
accretion, cooling and mass-loss by evolved stars is ignored. We thus
interpret the low-luminosity AGN in the AGN+SF phase as the culprit
responsible for the destruction of the gas reservoir. It is
possible that the feedback work of the AGN might have begun in an
obscured fashion during the SF phase, during which the feedback from
stars might have aided the gas removal.

Second, we perform the same analysis for each galaxy assuming a
measured line width of $70~ \rm kms^{-1}$ for the non-detected
galaxies. We find the same trend: starforming early-type galaxies have
substantial molecular gas reservoirs. A few AGN+SF early-types are
detected with diminished gas reservoirs, while others are undetected,
with gas masses inferred to be at least an order of magnitude
lower. The individual galaxy data points show that the depletion of
the molecular gas reservoir occurs within the AGN+SF phase about $200$
Myr after the start of star formation. The Seyfert AGN are all
undetected. For the individual galaxies, we perform a Bayesian
significance test and find significant evidence for populations of
galaxies with star formation more recent than 200 Myr and older than
this value having different molecular gas masses.

We then discuss the implications of our result and how it impacts our
understanding of AGN feedback. The process we see occurring suppresses
residual star formation in intermediate-mass, low redshift galaxies
and so could be termed a `suppression mode' of the AGN, which
maintains the red colors of early-type galaxies by suppressing
residual star formation. This mode stands in contrast to the
`truncation mode' which is thought to shut down star formation during
the formation of massive galaxies at high redshift.

The AGN phase responsible for this suppression mode is a
low-luminosity event and not a powerful quasar or radio galaxy
phase. The fact that we observe low-luminosity AGN impacting the fuel
for star formation in early-type galaxies is an important step towards
understanding galaxy formation.

\acknowledgements This work is based on observations made with the
IRAM 30 m telescope at Pico Veleta, near Granada, Spain. We thank the
anonymous referee for numerous suggestions and improvements to this
work. KS is supported by the Henry Skynner Fellowship at Balliol
College, Oxford. CJL acknowledges support from the STFC Science in
Society Programme. SK acknowledges a Leverhulme Early-Career
Fellowship, a BIPAC Fellowship at Oxford and a Junior Research
Fellowship from Worcester College, Oxford.

This work was supported by grant
No.~R01-2006-000-10716-0 from the Basic Research Program of the Korea
Science and Engineering Foundation to SKY. CM is a Marie-Curie
Excellence Team Leader and acknowledges grant MEXT-CT-2006-042754 of
the European Community. This research has made use of NASA's
Astrophysics Data System. \\

{\it Facilities:} \facility{IRAM:30m (A100, B100, A230, B230)}

\bibliographystyle{astroads}

\end{document}